\documentstyle[12pt]{article}
\catcode`@=11   % WATCH OUT !!!!! This must be reset in the end !!!!!
\def\IR{\relax\leavevmode{\rm I\kern-.18em R}}
\def\ZZ{\relax\leavevmode
       \ifmmode\mathchoice
       {\hbox{\sf Z\kern-.4em Z}}
       {\hbox{\sf Z\kern-.4em Z}}
       {\lower.9pt\hbox{\scriptsize\sf Z\kern-.36em Z}}
       {\lower1.2pt\hbox{\tiny\sf Z\kern-.36em Z}}
       \else{\sf Z\kern-.4em Z}\fi}
\def\RR{\relax\leavevmode
       \ifmmode\mathchoice
       {\hbox{\sf R\kern-.4em R}}
       {\hbox{\sf R\kern-.4em R}}
       {\lower.9pt\hbox{\scriptsize\sf R\kern-.36em R}}
       {\lower1.2pt\hbox{\tiny\sf R\kern-.36em R}}
       \else{\sf R\kern-.4em R}\fi}

\def\resetby#1#2{\@addtoreset{#2}{#1}}
\def\seceq{\@addtoreset{equation}{section}
              \def\theequation{\thesection.\arabic{equation}}}

\def\Label#1{\label{#1}%
                \smash{\hbox to0pt{\raise1ex\hbox{\tiny[#1]}\hss}}}
\def\noLabels{\let\Label=\label}

\thicklines     

\reversemarginpar
\catcode`@=12

\begin{document}

\begin{titlepage}

\begin{flushright}

VPI-IPPAP-03-10\\

hep-th/0309239\\

\end{flushright}

\begin{center}

\vskip 1cm

{\large \bf
          {A GENERAL THEORY OF QUANTUM RELATIVITY}}\\[5mm]
{\bf Djordje~Minic\footnote{e-mail: dminic@vt.edu}
and Chia-Hsiung Tze\footnote{e-mail: kahong@vt.edu}} \\[3mm]
            {\it Institute for Particle Physics and Astrophysics}\\
            {\it Department of Physics}\\
            {\it Virginia Tech}\\
            {\it Blacksburg, VA 24061}\\[5mm]

\vskip 1cm

{\bf ABSTRACT}\\[3mm]

\parbox{4.9in}{The geometric form of standard quantum mechanics is compatible with the two postulates: 1) The laws of physics are invariant under the choice of experimental setup 
and 2) Every quantum observation or event is intrinsically statistical.  These postulates remain compatible within
a background independent extension of quantum theory with a local intrinsic time implying the relativity of the concept of a quantum event.  In this extension the space of quantum events becomes dynamical
and only individual quantum events make sense
observationally.
At the core of such a general theory of quantum relativity is the three-way interplay between  the
symplectic form, the dynamical metric
and non-integrable 
almost complex structure of the space of quantum events. Such a formulation provides a missing conceptual ingredient in
the search for a background independent quantum theory of gravity and matter.
The crucial new technical element
in our scheme derives from a set
of recent mathematical results on certain infinite dimensional almost Kahler manifolds which replace the complex projective spaces of standard
quantum mechanics.}

\end{center}

\end{titlepage}

In this letter, as a stepping stone to its possible extensions, standard quantum mechanics (QM) is recast in the language of complex geometry by way of only two compatible postulates. The latter show that, just as thermodynamics, special relativity and general relativity (GR), QM, in spite of appearance, does belong, in Einstein's categorization, to ``theories of principles'' \cite{stachel}.  These two postulates to be stated below form a physically more intuitive rendition of the mathematical axioms of Landsman \cite{landsman}.  They make manifest the very rigid  structure
of the underlying state space (the space of
quantum events), the complex projective space $CP(n)$.  As such they also
underscore the relational \cite{qrel} and information theoretic
nature of quantum theory \cite{wootters}.  Most importantly, this perspective points to a possible extension of QM along the line discussed in \cite{tzeminic}, one relevant to a background independent formulation of quantum gravity.  Such a generalization is achieved, in analogy to what is done with the spacetime structure in GR, in a twofold way 1) by relaxing the integrable complex structure of the space of events and 2) by making this very space of events (that is, its metric and symplectic
and therefore its almost complex structure), the arena of  quantum dynamics, into a dynamical entity.
One of the byproducts of such an extended quantum theory is the notion of an intrinsic, probabilistic local time.  This quantum time is rooted in the strictly almost Kahler geometry of a dynamically evolving, diffeomorphism invariant state space of events. In physical terms, this nonintegrable almost complex structure implies a relaxation of the absolute global time of QM to an intrinsic, relative, local time. This novel feature is, in our view, the key missing
conceptual
ingredient in the usual approaches to the background
independent formulation of quantum gravity. 
The main technical thrust
of the present paper is contained in a set
of very recent mathematical results of Haller and Vizman \cite{vizman} concerning the
category of infinite dimensional almost Kahler manifolds which, in our view, naturally replaces
the category of complex projective spaces of standard QM. These results enable us
to significantly sharpen the geometric formulation of our previous more heuristic proposal \cite{tzeminic}.

First we recall that, among their many available formulations, the axioms of standard QM \cite{auletta} can take
a very elegant, simple C*-algebraic form  \cite{landsman}.  The Landsman formulation offers an $unified$ view of both quantum and classical mechanics thereby suggesting its structural closeness to geometric QM  \cite{geomqm}. We recall that in the latter setting, a quantum system is described by an infinite classical Hamiltonian system, albeit one with very specific Kahler constraints.  Here we seize on this formal closeness by providing the physical, geometric counterparts  of these Landsman
axioms. 
Paraphrasing \cite{landsman}, the first axiom states that the space of pure states is a Poisson space with
a transition probability. More precisely his defining Poisson bracket is exactly that in geometric QM \cite{geomqm}. Then, as detailed in Landsman's book  \cite{landsman}, the first axiom says that the essential physical information is carried by a
well defined symplectic (i.e. a non-degenerate symplectic 2-form) and metrical structures on the space of states.
The second axiom further specifies the transition probability to be that of  standard
QM, namely the metric information of
the Cayley-Fubini-Study type \cite{geomqm}, the natural, unique metric on $CP(n)$.
(The third axiom deals with superselection sectors, which, for simplicity are hereby ignored.). 
It suffices to say that Landsman's axioms can be shown \cite{landsman} to imply the usual geometric structure of QM, in particular the uniqueness of $CP(n)$ as the space of pure states. 

 Moreover the Landsman axioms as translated above can be understood in the following physically more intuitive manner.
To do so, we first recall Bohr's dictum that
``(quantum) physical phenomena are observed {\it relative} to different
experimental setups'' \cite{jammer}. This statement
closely parallels the role that inertial reference frames
play in relativity theory. More accurately,  as paraphrased by Jammer  \cite{jammer}, this viewpoint reads:
``...just as the choice of a different frame of reference in relativity affects the result of a
particular measurement, so also in quantum mechanics the choice of a different
experimental setup has its effect on measurements, for it
determines what is measurable''.  Thus while the observer does choose what to observe by way of a particular experimental setup, he or she cannot influence quantitatively the measured value of a particular observable.
Thus in analogy with the postulates of special
relativity and in the place of Landsman's axioms we propose the following two quantum postulates:

I) The laws of physics are invariant under the 
choice of the experimental set up.
Mathematically, we thus prescribe that, as in classical mechanics, there is a
well defined symplectic structure which stands for the
classical kinematical features of the measurement process. 

To expand on this Postulate 1, we should first note that, in a broader setting, it actually allows for a general Poisson structure. However, by confining for  simplicity, to a theory with no selection rules, we thus restrict ourselves to a symplectic structure. Now the classical symplectic structure is an inherent property that comes with the measurement
device whose readings are then statistically analyzed in the sense of statistical inference
theory. That a measurement device always comes together with a symplectic structure can be seen as follows: Take a system on which we perform physical measurements, it is described by
a certain Hamiltonian (or Lagrangian) so that the classical dynamics can be well defined.  Consider
a coupling of this system to another one, a measurement device, so that both the interaction
Hamiltonian and the Hamiltonian of the measurement device are in principle known.
(This is the classic set-up considered for example in the literature on decoherence
\cite{decoh}.) The measurement process is then in principle described by the interaction part of the total
Hamiltonian. Knowing the Hamiltonian assumes knowledge of a 
well defined symplectic structure.
Thus the existence of a symplectic structure is an intrinsic property that
comes with a measuring set-up.

So the first postulate asserts the existence of
a natural classical closed symplectic 2-form $\Omega$, 
\begin{equation}
d\Omega=0.
\end{equation}
Namely  the state space is an even dimensional symplectic Poisson manifold.
This is the mathematical rendition of our postulate I.

Next we make a principle out of another dictum of Bohr (and of Heisenberg
and Pauli) on the existence of $primary$ probabilities in nature:

II) Every quantum observation (reading of a given measurement device) or quantum event, is irreducibly
statistical in nature. These events, being distinguishable by measurements, form points of a statistical (informational) metric space. There is then a natural, unique statistical distance function
on this space of quantum events, the famous
Fisher distance \cite{wootters} of statistical inference theory \cite{arima}.

More precisely, from the seminal work of Wootters \cite{wootters}, 
a natural statistical distance on the
space of quantum events
is uniquely determined by 
the size of statistical fluctuations occuring in
measurements performed to tell one event from another. 
This distance between two statistical events is given in terms of the number of distinguishable
events, thus forming a space with the  the associated Riemannian  metric $ds^2 \equiv  \sum_i \frac{dp_i^2}{p_i} = \sum dX_i^2$ where
$p_i \equiv X_i^2 $ denote individual probabilities.
This distance in the probability space is nothing but the celebrated Fisher distance of information theory 
and can be rewritten as \cite{wootters}
\begin{equation}
ds_{12} = cos^{-1}(\sum_i \sqrt{p_{1i}} \sqrt{p_{2i}}).
\end{equation}
This is the mathematical content of our second postulate.

By introducing the coordinate representation for the square roots of probabilities
in the even dimensional symplectic manifold we can rewrite this expression as 
the Cayley-Fubini-Study metric of QM, using the standard notation, 
$
ds_{12}^2=4(cos^{-1}|\langle \psi_1|\psi_2\rangle|)^2 = 4(1 -
|\langle \psi_1|\psi_2 \rangle|^2)\equiv 4(\langle d\psi|d\psi\rangle
- \langle d\psi|\psi\rangle\langle \psi|d\psi\rangle),
$
albeit up to a multiplicative constant factor.
As emphasized by Wootters \cite{wootters}, the
statistical Fisher metric, does not have 
a priori anything to do with a metric on
the projective complex Hilbert space of QM.  Yet this information theoretic formulation of the metrical structure of QM, upon folding in the above compatible symplectic structure (as required by the first postulate), tells us that these distances are in fact one and the same; so we identify the multiplicative constant to be the Planck constant $\hbar$. This established the physical meaning of $\hbar$.
In other words, the measuring device that provides readings of distinguishable statistical events
has to be such that the physics is left invariant under the changes of basis on
this even dimensional space of square roots of the probabilities, and is
compatible with the macroscopic changes of basis in the sense of classical Hamiltonian
dynamics. Note that by expressing the statistical metric as a function of
square roots of the probabilities on the even dimensional space of quantum events,
we immediately saturate the Born rule $\sum_i p_i \equiv \sum_i X_i^2 =1$.

To have QM from the above, the crucial element is the {\it compatibility} condition between the 
symplectic ($\Omega$, from our postulate I) 
and metric ($g$, from our postulate II) structures. Such a condition
implies the existence of the $integrable$ complex structure
(in the same matrix notation) \cite{geomqm}
\begin{equation}
J=g^{-1}\Omega, \quad J^2 =-1.
\end{equation}
Our postulates I and II are just a physical rendition of Landsman's axioms.  As such they imply that the state space of QM is $CP(n)\equiv U(n+1)/U(n) \times U(1)$ \cite{landsman}, $n+1$ being the dimensionality of the Hilbert space. So statements about quantum mechanics are simply statements about the
geometry of complex projective spaces \cite{geomqm}.
Observe that in the classical limit (formally defined by taking $\hbar \to 0$)
the metrical information is lost, and only the classical symplectic information
is retained, in accordance with Landsman's axiomatic approach
\cite{landsman}.

Finally, we note that the above two postulates still leave free the choice of a Hamiltonian whose dynamics could thus be either nonlinear or nonunitary. For example one could have a Kibble-Weinberg
non-linear quantum dynamics \cite{geomqm}.  Only the condition of democracy among all observables  i.e. that the energy 
should not be different from any other observable, namely that the Hamiltonian evolution along the $CP(n)$ be also a Killing flow,
picks out the standard linear unitary evolution.
(In this respect the superposition principle is crucially related to the
geometric fact that $CP(n)$ can be viewed as a space of complex lines passing
through the origin of a complex space.)
\footnote{Quantum entanglements are entirely determined by the
unique geometric properties of complex projective spaces.
Tensoring of complex Hilbert spaces implies tensoring of
complex projective spaces the product of which 
can be embedded in a higher dimensional complex
projective space \cite{geomqm}.}

In light of our foregoing postulates, one might wonder to what extent 
the structure of quantum mechanics is actually fixed. In other words, how much can one tamper with standard QM given the rigid geometry of complex projective spaces? While it is true that many different axiomatic systems imply standard QM, the latter may appear more yielding to and suggestive of possible deformation(s) in one formulation than in another.  So in a more positive and specific vein, we may ask: how QM, seen as a contingent theory in the very perspective of our postulates, can be modified or deformed so as to still be compatible with the two principles stated above?  

Before putting forth a concrete answer, we briefly circumscribe this extension issue by summing up some aspects of the strikingly robust mathematical rigidity of standard QM, of its apparent stability against several deformations of the state space which naturally come to mind.
(See also the discussion in
\cite{geomqm}, especially the works of Ashtekar and Schilling and of Gibbons).

Thus let us replace $CP(n)$ by a projective Kahler manifold $M$; the projective and complex structures being shown above to be the key features worth preserving.  Let $M$ be homogeneous and isotropic.  Since the latter two properties are known to be necessary (see Hatakeyama \cite{geometry})
for the classical phase space dynamics, it is sensible to assume the same for 
the quantum state space of which the classical phase space is a reduced subspace.  
Furthermore we allow $M$ to be of constant positive sectional curvature since the latter governs the finiteness and sign of Planck's constant.  Finally, for simplicity and barring global phenomena, assume $M$ to be connected and simply- connected.  Then there is, blocking our path, a fundamental theorem of Hawley \cite{geometry} and Igusa \cite{geometry}.  It states that, for finite $n$, the projective spaces are up to isomorphisms the only connected, simply connected and complete Kahler manifolds of constant and positive holomorphic sectional curvature, namely they are isomorphic to $CP(n)$.  Moreover, a recent very strong result 
of Siu and Yau and of Mori \cite{geometry} shows that the requirement of positive bisectional curvature $alone$ necessarily implies that the underlying manifold is $CP(n)$.   Whether the above stringent theorem extends to the infinite dimensional case is, to our best knowledge, not known. The likelihood of an affirmative answer to that question is, in our view, strongly hinted by a theorem of Bessega 
\cite{geometry}, namely that every infinite dimensional Hilbert space is diffeomorphic with its unit sphere.  If so, there is no other infinite dimensional connected, simply connected, homogeneous and isotropic Kahler manifold beside $CP(\infty)$. Contrasting with the arbitrariness in the topology and geometry of the classical phase space and its symplectic structure, is the striking universality of the $CP(n)$ of QM, where the metric, symplectic and complex structures are so closely interlocked that the only freedom left is the values of n and  $\hbar$.
Exploring another way to alter the kinematics, we may ask: How about taking a Kahler manifold with simply a constant scalar curvature for a space of states? Here by a theorem of Yano and Kon \cite{geometry} such a manifold is necessarily flat.  What if we seek  a space which is a small deformation of a Kahler manifold?  Here yet another of Yano and Kon's theorems \cite{geometry}, holding in finite dimensions, asserts that, except in dimension six, there is NO nearly Kahler manifold. We must also stress that the state space $M$ must be a projective space if one insists that the observables be a sufficiently wide set and they close on an associative algebra \cite{geomqm}. In fact it is known that for the set of observables 
to be {\it maximal}, $M$ must be a manifold of constant 
holomorphic sectional curvature, having the maximal number of Killing vectors (as discussed by Ashtekar and Schilling 
in \cite{geomqm}). 

For completeness and in view of what follows, we may look at the extension problem from yet another broader angle: We observe that, besides the $n$-spheres, which except for $n=2$, are neither complex nor projective, the real projective spaces $RP(n)$ and $CP(n)$ of standard QM, there are the quaternionic projective spaces $HP(n)= Sp(n+1)/Sp(n) \times Sp(1)$ and the sixteen dimensional octonionic projective Cayley-Moufang Plane $CaP(2)= F_4/SO(9)$. Most remarkably the spaces listed above are all orbit spaces of the orthogonal, unitary, symplectic and exceptional groups, respectively. They are not only projective spaces but also Cartan's symmetric compact spaces of rank one (CROSSes). Symmetric spaces have the defining property that their curvature is invariant under parallel transport. Notably all the geodesics of CROSSes, in their canonical metrics, are simple, closed (periodic) and of a common length, as
discussed by Berger \cite{geometry}. It is in fact this feature, rather than linearity, which underlies the possibility of superposition in QM \cite{geomqm}.
The natural metric on the above projective spaces is of the Fubini-Study (FS) type, taken over the
appropriate number fields, the complex numbers, quaternions  and octonions.\footnote{In the case of the Cayley-Moufang plane the natural metric is encoded in the data
of the Jordan algebra of 3 by 3 hermitian matrices over octonions \cite{mostow}, whose
automorphism group is the exceptional group $F_4$.}   As candidate state spaces these spaces do give consistent quaternionic and octonionic quantum mechanics which are extensions of standard complex QM  \cite{qoct}.  However, with their unique FS type metrics, they have their own rigidity, are thus not  the sought after deformations in our purported search for a background independent complex quantum mechanics.

Coming back to the inflexible kinematics of standard complex QM where the space of quantum events must be $CP(n)$, we do see a new physically revealing way to relax its rigidity.  Our new angle of attack rests with the observation that the rigid geometry of $CP(n)$ reflects the very striking interplay of a triad of structures connected with $g$, $\Omega$ and $J$ within the state space. They are its  Riemannian structure with its generic holonomy or stabilizer group $O(2n) $, its symplectic structure with its stabilizer group $Sp(4n,R)$ and its complex 
structure $J$ with its group $GL(n,C)$.  The intersection of the three associated Lie groups results in a subgroup of $O(2n)$, the unitary group $U(n,C)$, hence in the unitarity in QM, the hermiticity of the observables and the Hermitian geometry of $CP(n)$.  Notably, any two elements of this triad plus their mutual compatibility condition imply the third. 

The physical basis of the above triadic linkage is lucidly discussed by Gibbons and Pohle in  \cite{geomqm}.  Namely,  observables  in QM play a $dual$ role as: 1) providers of outcomes of measurements  and 2) generators of canonical transformations.  Indeed the almost complex structure $J$ is nothing but the dual representative of the measurable observable $g$, as it generates canonical transformations corresponding to this metric, particularly time evolution.  Thereby the time $t$ in QM is connected in a one to one way to the FS metric and hence to the almost complex structure $J$. This connection is explicit in the Aharonov-Anandan relation $\hbar ds = 2\Delta (H)  dt$, where $\Delta(H)^2 \equiv  < H^2> - <H>^2$ is the uncertainty in the energy (see Anandan in  \cite{geomqm}). This linear relation between the metric and time shows the probabilistic nature of time and time as a correlator between statistical
distances measured by different systems.  Now if an almost complex structure is given on a manifold $M$, it does not yet follow that $M$ is complex, namely that a complex coordinate system can be $globally$ introduced on $M$. For that to be the case, the almost complex structure (ACS)  must be $integrable$. It suffices to state that, by the Newlander-Nirenberg theorem \cite{nn}, the necessary and sufficient condition for integrability or ``globality'' of the local almost complex structure is given by the vanishing of the Nijenhuis torsion tensor.  Note that $J$ $is$ integrable on $CP(n)$ for any $n$; there lies the $absolute$ (Galilean like) $global$ time in standard QM.  Physically then, an ACS on a state space which fails to be integrable  means from our former argument, that the corresponding quantum theory no longer has a global time but rather a local, relational time.  This more provincial, local notion of time is in better accord with our expectations from GR.  As a result there is a relativity among observers of the very notion of a quantum event\footnote{This possibility was also discussed in \cite{isidro}.}.

Our principles (I) and (II) as stated above clearly display
quantum theory as what might be called a
special theory of quantum relativity. It is then only natural to take the next logical step,  to go beyond and formulate
a general theory of quantum relativity.  This extension is accomplished
by allowing both the metric and symplectic form
on the space of quantum events to be no longer rigid but fully
dynamical entities.
In this process, just as in the case of spacetime in GR,  the space of quantum events becomes dynamical
and only individual quantum events make sense
observationally. 

Specifically, we do so by relaxing our postulate II to allow for {\it any} statistical (information) metric all the while
insisting on the compatibility of this metric with the symplectic structure
underlying our postulate I. Physics is therefore required to be diffeomorphism
invariant in the sense of information geometry \cite{arima} such
that the information geometric and symplectic structures remain compatible, requiring only a $strictly$ ( i.e non-integrable) almost complex structure $J$.
Once we relax postulate II, so that any information metric is allowed,
the relativity of canonical quantum mechanical experiments (such as the double-slit
experiment) becomes possible and would provide an experimental test of our proposal.

Our extended framework readily implies that the wave functions labeling the event space, while still unobservable, are no longer relevant. They are in fact as meaningless as coordinates
in GR.  There are no longer issues related to reductions of wave packets and associated measurement
problems. At the basic level of our scheme, there are only dynamical correlations of quantum events.
From the previous analysis and in the spirit of constructing an ab initio quantum theory (of principles) of matter and gravity, we can enumerate the main structural  features one may want in such a scheme for the space of quantum events, that  1) it has a symplectic structure 2) it is strictly almost Kahler  3) it is the base space of a $U(1)$ bundle and 4) it is diffeomorphism invariant.
We recall that the state space  $CP(\infty$)  is a linear Grassmannian manifold, $CP(n)$ being the space of complex lines in $C^{n+1}$ passing through the origin. We seek
a coset of $Diff(C^{n+1})$ such that locally looks like $CP(n)$ and allows
for a compatibility of the metric and symplectic structures, expressed in
the existence of a (generally non-integrable) almost complex structure.

The following nonlinear Grassmannian\footnote{Note that this non-linear Grassmannian, is 
{\it different} from the manifold
$Diff(\infty, C)/(Diff(\infty-1, C) \times Diff(1,C)$
we have proposed in our previous paper \cite{tzeminic}. It appears that
the basic geometric properties, such as the possible existence of an almost complex
structure, of 
$Diff(\infty, C)/(Diff(\infty-1, C) \times Diff(1,C)$ are not known. We thank
C. Vizman for correspondence on this issue. We also thank A.~Ashtekar and R.~Penrose
for insisting on the existence of an almost complex structure on the generalized
space of states.} 
\begin{equation}
Gr(C^{n+1}) = Diff(C^{n+1})/Diff(C^{n+1},C^n \times \{0\}) ,
\end{equation}
with $n = \infty$ satisfies the above requirements, thus sharpening the
geometrical information of the proposal made in \cite{tzeminic}.

Indeed this infinite (even for finite $n$) dimensional space $Gr(C^{n+1})$ is modeled on a Frechet space. Very recently, its study was initiated by Haller and Vizman \cite{vizman}.   
Firstly it is a nonlinear analog of a complex Grassmannian since it is the space of (real) co-dimension 2 submanifolds, namely a hyperplane $C^n \times [0]$ passing through the origin in $C^{n+1}$.  Its holonomy group $Diff(C^{n+1} , C^n\times \{0\})$ is the group of diffeomorphisms preserving the hyperplane  $C^n \times \{0\}$ in $C^{n+1}$. Just as $CP(n)$ is a coadjoint orbit of $U(n+1)$, $Gr(C^{n+1})$ is a coadjoint orbit of the group of volume preserving diffeomorphisms of $C^{n+1}$. As such it is a symplectic manifold with a canonical Kirillov-Kostant-Souriau symplectic two form $\Omega$ which is closed ($d\Omega=0$) but not exact. Indeed the latter 2-form integrated over the submanifold is nonzero; its de Rham cohomology class is integral.  This means that there is a principal 1-sphere, a $U(1)$ or line bundle over $Gr(C^{n+1})$ with curvature $\Omega$. This is the counterpart of the 
$U(1)$-bundle of $S^{2n+1}$ over $CP(n)$ of quantum mechanics. It is also known that there is an almost complex structure given by a 90 degree rotation in the two dimensional normal bundle to the submanifold.  While $CP(n)$ has an integrable almost complex structure and is therefore a complex manifold, in fact  a Kahler manifold, this is $not$ the case with $Gr(C^{n+1})$. Its almost complex structure $J$ is by a theorem of Lempert \cite{lempert} strictly $not$ integrable in spite of its formally vanishing Nijenhius tensor. While the vanishing of the latter implies integrability in the finite dimensional case, one can no longer draw such a conclusion in the infinite dimensional Frechet space setting. However what we do have in $Gr(C^{n+1})$ is a strictly (i.e. non-Kahler) almost Kahler  manifold \cite{almost} since there is by way of the almost structure $J$ a compatibility between the closed symplectic 2-form $\Omega$ and the Riemannian metric $g$ which $locally$ is given by $g^{-1}\Omega = J$.\footnote{It would be very interesting to understand how unique is the structure
of $Gr(C^{n+1})$.}

Next, just as in standard geometric QM, the probabilistic interpretation lies in the
definition of geodesic length on the new space of quantum states (events) 
as we have emphasized before \cite{tzeminic},
\cite{geomqm}.
Notably since $Gr( C^{n+1})$ is only a strictly almost complex, its $J$ is only locally complex. This fact translates into the existence of only local time and local metric on the
space of quantum events. 
As we have proposed in our previous work \cite{tzeminic},  the local temporal
evolution equation is a geodesic equation on the space of quantum events
\begin{equation}
{d u^a \over d\tau} + \Gamma^{a}_{bc} u^b u^c =
\frac{1}{2 E_p}Tr(H F^a_b) u^b
\end{equation}
where now $\tau$ is given through the metric
$\hbar d\tau = 2 E_p dt$, where $E_p$ is the Planck energy.
$\Gamma^{a}_{bc}$ is the affine connection associated with this general metric 
$g_{ab}$ and $F_{ab}$ is a general curvature 2-form in the holonomy gauge group $Diff(C^{n+1},C^n \times \{0\})$.\footnote{The fact that the generalized geometric phase
is in $Diff(C^{n+1},C^n \times \{0\})$ should be also possible to test experimentally.}
This geodesic equation follows from the conservation of the
energy-momentum tensor 
$
\nabla_a T^{ab} =0
$
with 
$
T_{ab} = Tr(F^{ac}g_{cd}F^{cb} -\frac{1}{4} g_{ab} F_{cd}F^{cd}
+ \frac{1}{2E_p}H u_a u_b).
$
Since both the metrical and symplectic data are also contained in
$H$ and are $\hbar \to 0$ limits of their quantum counterparts \cite{geomqm}, \cite{tzeminic}, we have here a non-linear ``bootstrap'' between the space of
quantum events and the dynamics.
The diffeomorphism invariance of the new phase space
is explicitly taken into account in the following dynamical scheme 
\cite{tzeminic}:
\begin{equation}
\label{BIQM1}
R_{ab} - \frac{1}{2} g_{ab} R  - \lambda g_{ab}= T_{ab}
\end{equation}
($\lambda = \frac{n+1}{\hbar}$ for $CP(n)$; in that case $E_p \to \infty$).
Moreover we demand for compatibility
\begin{equation}
\label{BIQM2}
\nabla_a F^{ab} = \frac{1}{E_p} H u^b.
\end{equation}
The last two equations imply via the Bianchi identity a conserved
energy-momentum tensor, $\nabla_a T^{ab} =0$ .  The latter, taken
together with the conserved ``current'' $j^b \equiv \frac{1}{2E_p} H u^b$,
i.e. $\nabla_a j^a =0$,
results in the generalized geodesic Schr\"{o}dinger equation.

These local dynamical equations are precisely the ones we have proposed in 
our previous paper
\cite{tzeminic}. The fact that the space of quantum events should be
$Gr(C^{n+1})$ sharpens the global geometric structure of our proposal.
As in GR it will be crucial to understand the global features of various
solutions to the above dynamical equations.

Finally, we have argued in the previous paper \cite{tzeminic}, that the
form of $H$ (the Matrix theory Hamiltonian in an arbitrary background), viewed as a ``charge''  may be determined in a quantum theory of
gravity by being encoded in the non-trivial topology of the space of quantum events. 
This may well be the case here with our non linear Grassmannian which is non-simply connected \cite{vizman}. However definite answers to this and many other  more concrete questions must wait  until  greater details are known on the topology and differential geometry (e.g. invariants, curvatures, geodesics) of $Gr(C^{n+1})$.  In the meantime we hope to have laid down here the conceptual and mathematical foundations of what may be called a general theory of quantum relativity.

{\bf Acknowledgments:}
We are specially indebted to a) Cornelia Vizman for invaluable correspondence and suggestions on the nonlinear Grassmannian manifold and b) the referee whose very perceptive critical remarks and suggestions have greatly enhanced the coherence of our paper.  We are also very
grateful to N. Landsman for pointing out and discussing his work with us.
We also wish to warmly thank A.~Ashtekar, C.~Bachas, V.~Balasubramanian, J.~de~Boer, R.~Dijkgraaf, L.~Freidel,
E. Gimon, M.~G\"{u}naydin, C.~Hull, V.~Jejjala, N.~Kaloper, K.~Krasnov, F.~Markopoulou, R.~Penrose, S.~Ross, C. Rovelli,
L.~Smolin, A.~Schwarz, H.~Verlinde, E.~Verlinde for useful discussions.
We would also like to acknowledge communications from P.~Freund, R.~G.~Leigh, Y.~Nambu, V.~P.~Nair, G.~Gibbons, C.~Rovelli
and T.~Takeuchi after the appearance of the first version of this paper.
D.M. would like to acknowledge the stimulating research environments of the Amsterdam workshop on string theory
and the Aspen Center for Physics where some of this work was done.

\end{document}